\documentclass[aps,twocolumn,superscriptaddress,pra]{revtex4}
\usepackage[T1]{fontenc}
\usepackage{amsmath,bbm} %per le formule matematiche
\usepackage{mathtools}
\usepackage{physics} %per la notazione di dirac
\usepackage{amsfonts} %per l'operatore identità \mathbb{I}
\usepackage{booktabs} %toprule tabelle
\usepackage{graphicx} %importazione immagini
\usepackage{float} %posizionamento H immagini
\usepackage{multirow}
\usepackage[caption=false]{subfig}
\usepackage{hyperref}
\usepackage{xcolor}
\usepackage{upgreek}
\usepackage{siunitx}
\usepackage{verbatim}
\usepackage{ulem}
\usepackage{times,mathptm}

\begin{document}

%\title{Un nuovo metodo per la stabilizzazione attiva tra seed e pompa in un optical parametric oscillator e applicazione nella generazione di stati squeezed}
\title{Novel technique for the active stabilization of the relative phase between seed and pump in an optical parametric oscillator}

\author{Simone~Cialdi}
 \email{simone.cialdi@mi.infn.it}
\affiliation{Dipartimento di Fisica ``Aldo Pontremoli'', Universit\`a degli Studi di Milano, via~Celoria~16, I-20133 Milan, Italy}
\affiliation{Istituto Nazionale di Fisica Nucleare, Sezione di Milano, I-20133 Milan, Italy}

\author{Edoardo~Suerra}
 \email{edoardo.suerra@unimi.it}
\affiliation{Dipartimento di Fisica ``Aldo Pontremoli'', Universit\`a degli Studi di Milano, via~Celoria~16, I-20133 Milan, Italy}
\affiliation{Istituto Nazionale di Fisica Nucleare, Sezione di Milano, I-20133 Milan, Italy}

\author{Matteo~G.~A.~Paris}
 \email{matteo.paris@fisica.unimi.it}
\affiliation{Dipartimento di Fisica ``Aldo Pontremoli'', Universit\`a degli Studi di Milano, via~Celoria~16, I-20133 Milan, Italy}
\affiliation{Istituto Nazionale di Fisica Nucleare, Sezione di Milano, I-20133 Milan, Italy}

\author{Stefano~Olivares}
\email{stefano.olivares@fisica.unimi.it}
\affiliation{Dipartimento di Fisica ``Aldo Pontremoli'', Universit\`a degli Studi di Milano, via~Celoria~16, I-20133 Milan, Italy}
\affiliation{Istituto Nazionale di Fisica Nucleare, Sezione di Milano, I-20133 Milan, Italy}

\begin{abstract}
We design and demonstrate a novel technique for the active stabilization of the relative phase between seed and pump in an optical parametric oscillator (OPO). We show that two error signals for the stabilization of the OPO frequency, based on Pound-Drever-Hall (PDH), and of the seed-pump relative phase can be obtained just from the reflected beam of the OPO cavity, without the necessity of two different modulation and demodulation stages.
We also analyze the effect of the pump in the cavity stabilization for different seed-pump relative phase configurations, resulting in an offset in the PDH error signal, which has to be compensated. Finally, an application of our technique in the reliable generation of squeezed coherent states is presented.
\end{abstract}

\maketitle

\section{Introduction}
``Squeezing'' is a fundamental resource in continuous-variable quantum information science \label{GQI} and in gravitational-wave detection \cite{lough}. For these and other applications a key ingredient is the use of squeezed states, which are commonly generated via an optical parametric oscillator (OPO) \cite{breite,dauria09,buono12,porto,mandarino}. The basic elements of an OPO are an optical cavity (the resonator) with a nonlinear crystal. Whereas the cavity allows to select a particular frequency of the field, the crystal, suitably ``pumped'' by an input laser beam (the ``pump''), provides the ``squeezing''. The generated states can be further manipulated by changing their coherent amplitude and phase. However, this requires using a ``seed'', namely, a coherent state with a well-determined complex amplitude, that interacts with the non-linear crystal, leading to the so-called squeezed coherent states. Mechanical vibrations largely affect this kind of devices, forcing the use of an active stabilization. Different techniques have been developed to this aim: the Pound-Drever-Hall (PDH) technique \cite{pdh}, the homodyne locking \cite{heurs}, the modulation free technique \cite{hansch}, the tilt locking \cite{robins}, just to mention the most commonly used. Moreover, also the relative phase between the seed and the pump used for squeezing has to be stabilized. This has been performed using modulation techniques such as in Ref.~\cite{bowen, sun}, or in GEO600 \cite{schnabel} or through the weak pump depletion (WPD) technique \cite{denker}.

Here we present the theoretical basis and the demonstration of an innovative technique for the seed-pump relative phase stabilization of a OPO with an application to the generation of squeezed coherent states. With respect to to modulation techniques \cite{bowen}, ours allows to obtain two different error signals directly from the reflected beam. The first signal is for the OPO frequency stabilization, and it is based on the PDH technique, but taking into account the presence of the pumped nonlinear crystal inside the cavity. In fact, since the standard PDH technique is based on an empty cavity, here we have first to derived the theoretical model to include the effect of the crystal. The other signal is for the seed-pump phase stabilization, and it does not require an additional modulation and demodulation stage. Thanks to this method, we do not need a detection system for the pump, that is instead necessary, for instance, in the WPD technique \cite{denker}. Remarkably, our approach allows us to retrieve relevant information about the dynamics of the error signal and to show how the pump affects the PDH error signal and how to correct this error signal.

The article is structured as follows. In Sect.~\ref{sec:theory} the technique is theoretically described taking into account the presence of the crystal inside the cavity. The setup for the implementation and the experimental results are shown in Sect.~\ref{sec:results}. Our pump-seed stabilization technique si applied  in Sect.~\ref{s:application} to generate squeezed choerent states. Finally, we draw some concluding remarks in Sect.~\ref{s:concl}.

\section{Theory}\label{sec:theory}
\begin{figure}[tb]
\centering
\includegraphics[width=0.9\columnwidth]{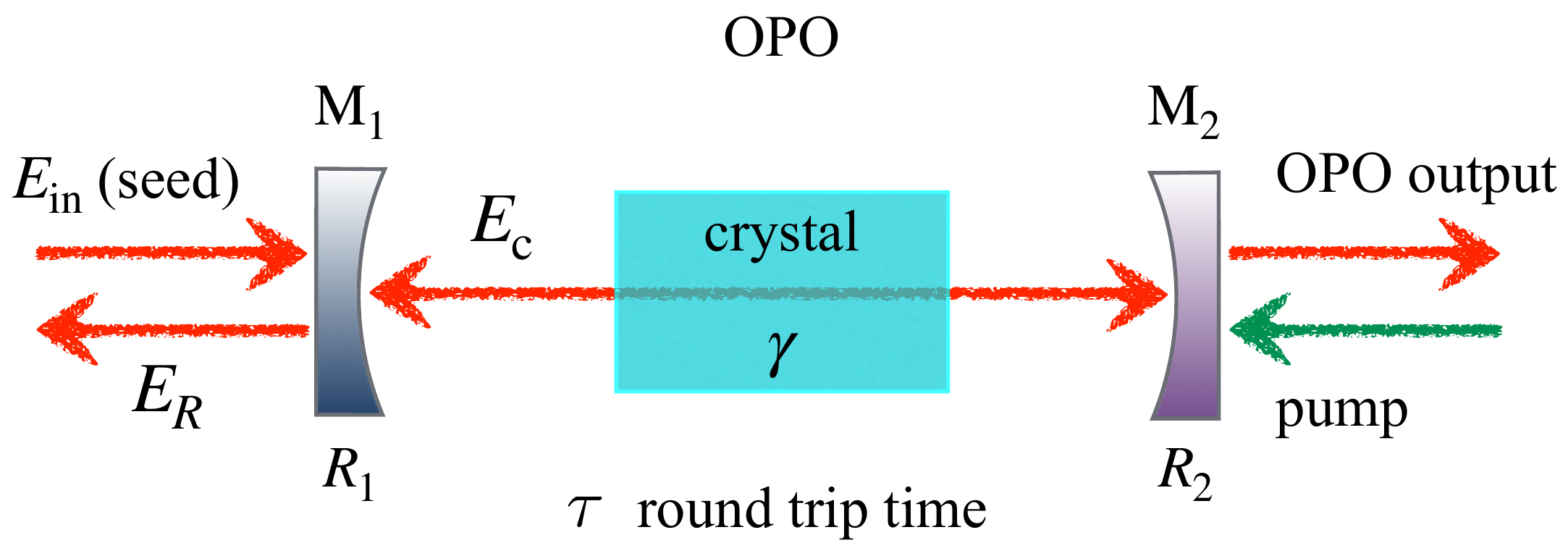}
\vspace{-0.2cm}
\caption{Scheme of the OPO with the main elements. See the text for details.
}
\label{fig:OPO}
\end{figure}
As in the case of the usual PDH stabilization technique, our approach is based on the detection of the beam reflected off the cavity but in the presence of a nonlinear crystal in it, that is also pumped to produce squeezing. It is worth noting that the seed (at 1064 nm in our experiment, see Sect.~\ref{sec:results}) and the pump (at 532 nm) have different frequencies. Therefore, in this Section we develop a proper theoretical model to describe a two-mirror cavity with a pumped nonlinear crystal inside, namely the OPO.

As sketched in Fig.~\ref{fig:OPO}, the mirror ${\rm M_1}$, with reflectivity $R_1$, serves as the input coupler for the seed $E_{\rm in}$, while the mirror ${\rm M_2}$, with reflectivity $R_2$, acts as the input coupler for the pump and as the output coupler for the OPO field. The aim is now to obtain the amplitude of the reflected field $E_R$ as a function of the cavity frequency $\nu$, of the crystal nonlinearity $\gamma$ and of the pump phase $\phi_{\rm p}$.

If the OPO does not contain the nonlinear crystal, the electric field in the cavity $E_{\rm c}(t+\tau)$ after a round trip with duration $\tau$ can be written as the sum of the field already circulating inside the cavity $E_{\rm c}(t)$ and the input field $E_{\rm in}$, that is:
\begin{equation}
E_{\rm c} \left( t + \tau\right) = \sqrt{1-R_1} \, E_{\rm in} + \sqrt{R} \, e^{i \phi} E_{\rm c}\left(t\right)
\label{eq:cavity_eq}
\end{equation}
where $R=R_1 R_2\left(1-\Delta\right)^2$, $\Delta$ is the internal power losses for the single roundtrip, and $\phi=2\pi\nu/\Gamma$. Here,
$\nu=\nu_{\rm in}-\nu_{\rm c}$ is the detuning between the input and the cavity frequencies and $\Gamma$ is the free spectral range of the cavity.

In the presence of non-linear crystal inside the cavity, Eq.~(\ref{eq:cavity_eq}) has to be modified adding a term accounting for interaction between the light and the crystal. The interaction can be effectively described by introducing a suitable complex parameter, $\gamma$, which depends on the pump field amplitude and its phase. Assuming the roundtrip much faster than the intra-cavity fields dynamics, we can expand $E_{\rm c} \left( t + \tau\right)$ up to the first order in $\tau$, and  Eq.~(\ref{eq:cavity_eq}) leads to
\begin{equation}
\frac{d E_{\rm c}}{dt}~ \tau =  \sqrt{1-R_1} E_{\rm in} + \left( \sqrt{R} e^{i \phi} -1 \right)  E_{\rm c} + i \gamma E_{\rm c}^*\,.
\label{eq:NLcavity_eq}
\end{equation}
The fact that here appears the complex conjugate of the cavity field, $E_{\rm c}^*$, directly follows from the energy conservation in the non-linear interaction \cite{yariv}.
Therefore, at the equilibrium we have
\begin{equation}
\left(1- \sqrt{R} e^{i \phi}  \right)  E_{\rm c} - i \gamma E_{\rm c}^* = \sqrt{1-R_1} E_{\rm in}
\label{eq:fondamentale_eq}
\end{equation}
Without loss of generality, from now on we can set $E_{\rm in}=1$ (or, equivalently, we are measuring the fields in unit of the input field) and, thus, the phases of the involved fields are relative to the phase of the input field.
From Eq.~(\ref{eq:fondamentale_eq}) we obtain
\begin{equation}
E_{\rm c} =  \sqrt{1-R_1} \frac{1- \sqrt{R} e^{-i \phi} +i \left|\gamma\right|e^{i \phi_{\rm p}}}{1+ R - 2 \sqrt{R} \cos\phi -\left| \gamma \right| ^2}
\label{eq:Ec_equation}
\end{equation}
where $\gamma=\left|\gamma\right|e^{i \phi_{\rm p}}$, $\phi_{\rm p}$ being the pump phase relative to the seed one. It is worth noting that $E_{\rm c}$ can be continuously modulated from an amplification regime ($\phi_{\rm p}=-\pi/2$) to a deamplification one ($\phi_{\rm p}=\pi/2$). As a matter of fact, if $\gamma \to 0$ we find the usual equation of a two-mirror cavity without the non-linear crystal, as one may expect.

The actual value of $\left|\gamma\right|$ can be obtained experimentally and we can relate it to a measurable quantity as follows. In fact one has
\begin{equation}
G = \frac{G_+}{G_-} =  \left| \frac{ 1+ \delta }{ 1 - \delta }\right|^2
\label{eq:gain}
\end{equation}
where $\delta=|\gamma|/(1-\sqrt{R}e^{i\phi})$ and $G_+$ and  $G_- $ are the power of $E_{\rm c}$ in the amplification and deamplification regime, respectively. Since the OPO output (see Fig.~\ref{fig:OPO}) is clearly proportional to $E_{\rm c}$, then $G$ can be measured by monitoring the output power in the two regimes.

Now, we turn our attention on the reflected beam $E_{R}$ (see Fig.~\ref{fig:OPO}), since it will be used to obtain two different error signals for the stabilization of the OPO frequency and the stabilization of the seed-pump relative phase. The reflected beam $E_R$ is the sum of the field directly reflected from the input coupler, with an additive $\pi$ phase, and the transmitted one through ${\rm M}_1$, thus it can be written as (we still set $E_{\rm in} = 1$):
\begin{equation}
E_R =  -\sqrt{R_1} + \frac{\sqrt{1-R_1}}{\sqrt{R_1}}
\left( 
\sqrt{R} e^{i \phi} E_{\rm c} + i|\gamma| e^{i \phi_{\rm p}} E_{\rm c}^*
\right)\,.
\label{eq:Er_equation}
\end{equation}
By substituting Eq.~(\ref{eq:Ec_equation}) in Eq.~(\ref{eq:Er_equation}), we explicitly find
\begin{align}
E_R &= \frac{\left(\sqrt{R} e^{i \phi} -R_1  \right) \left(1- \sqrt{R} e^{-i \phi}  \right)}
{\sqrt{R_1} \left( 1+ R - 2 \sqrt{R} \cos\phi  - | \gamma|^2 \right) }\nonumber\\
&\hspace{1.0cm}+\frac{ i \left[(1-R_1) e^{i\phi_{\rm p}} + | \gamma|\right] | \gamma|}
{\sqrt{R_1} \left(  1+ R - 2 \sqrt{R} \cos\phi  - | \gamma|^2 \right) }\,.
\label{eq:Er_final}
\end{align}
Again, if $\gamma \to 0$, then $E_R$ reduces to the usual reflected field without the non-linear crystal.
$E_R$ depends on the detuning between seed frequency and cavity frequency though $\phi$, but also on the pump phase $\phi_{\rm p}$.

\begin{figure}[tb]
\centering
\includegraphics[width=0.4\textwidth]{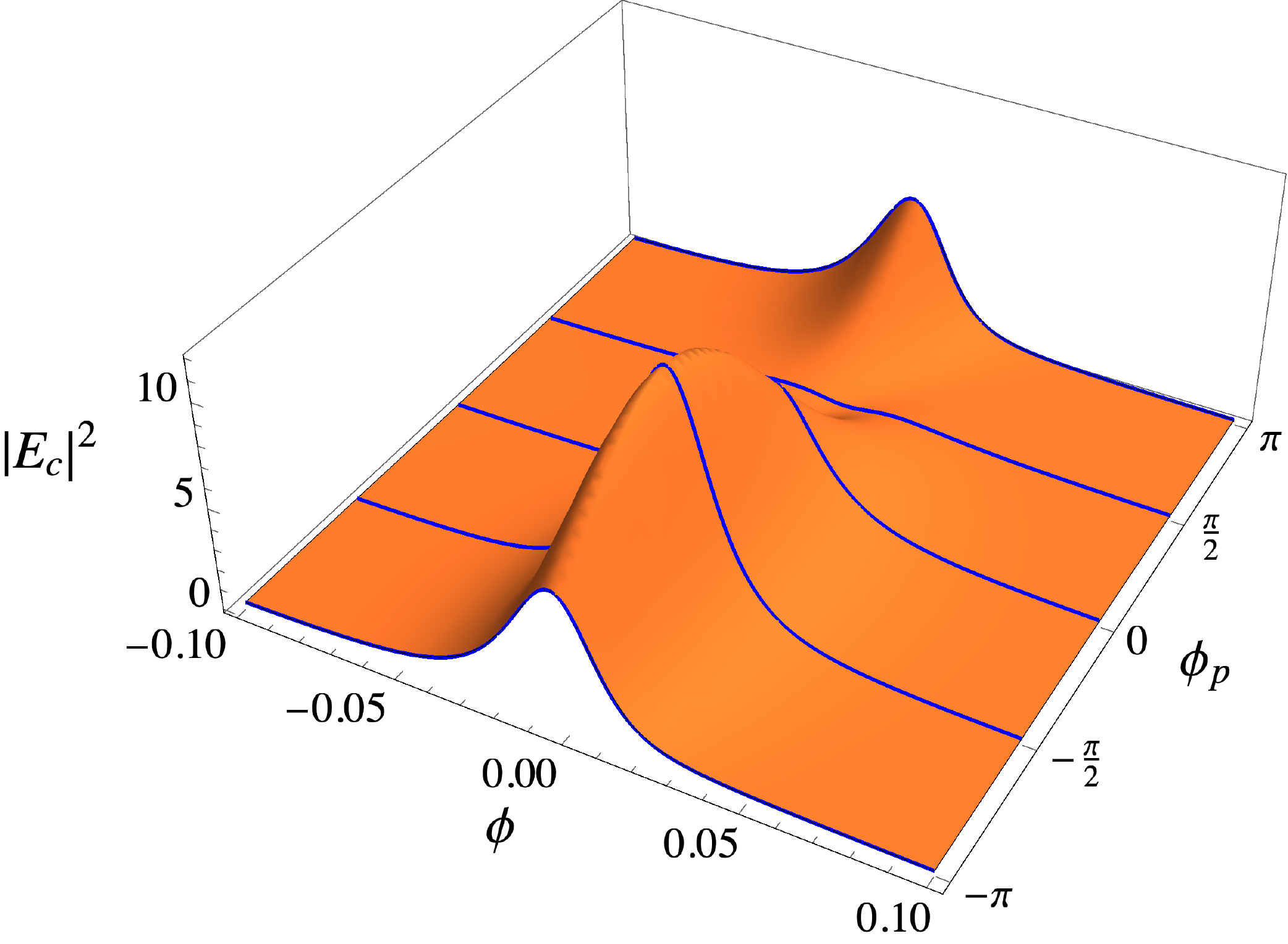}
\includegraphics[width=0.4\textwidth]{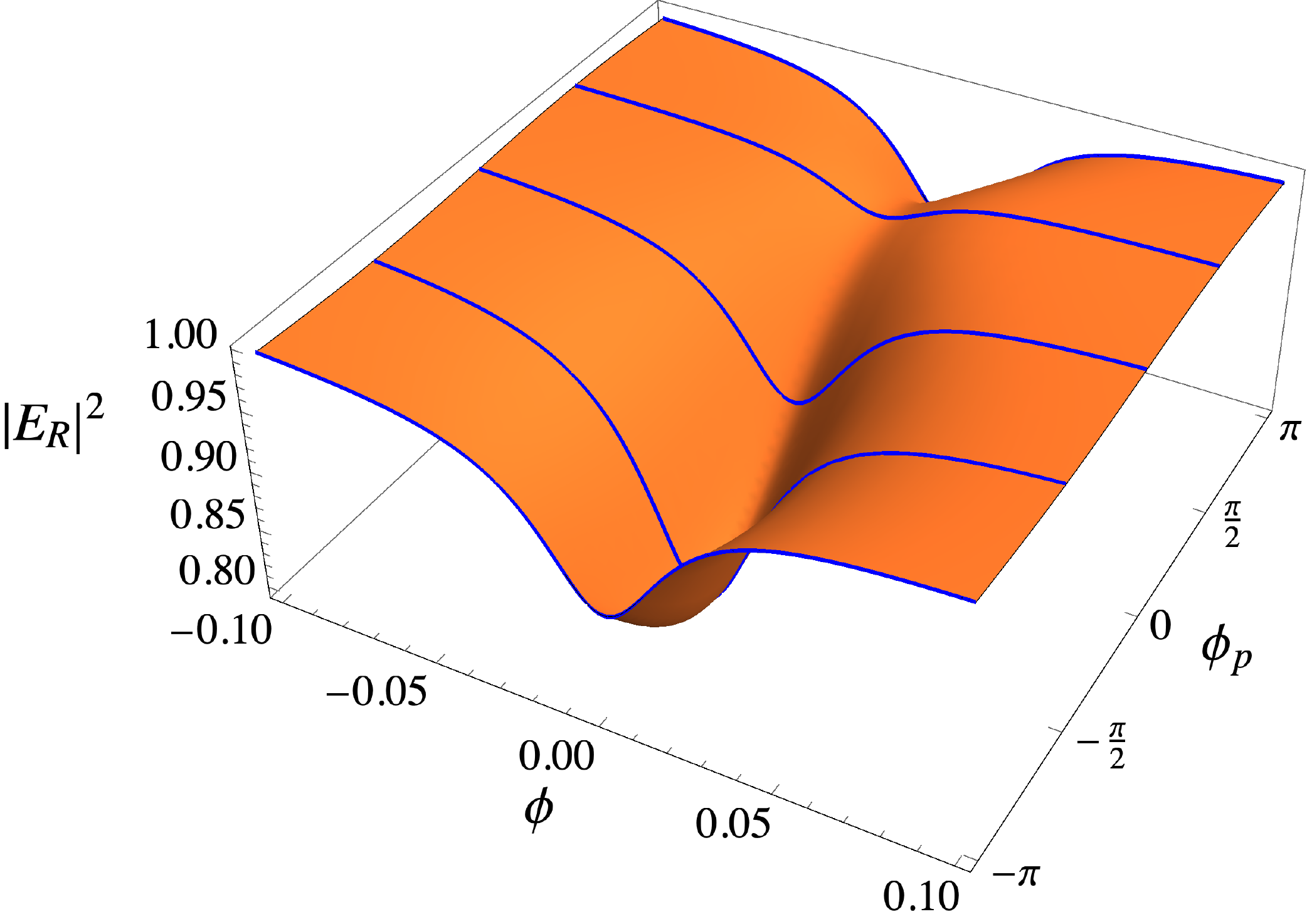}\\[2ex]
\includegraphics[width=0.4\textwidth]{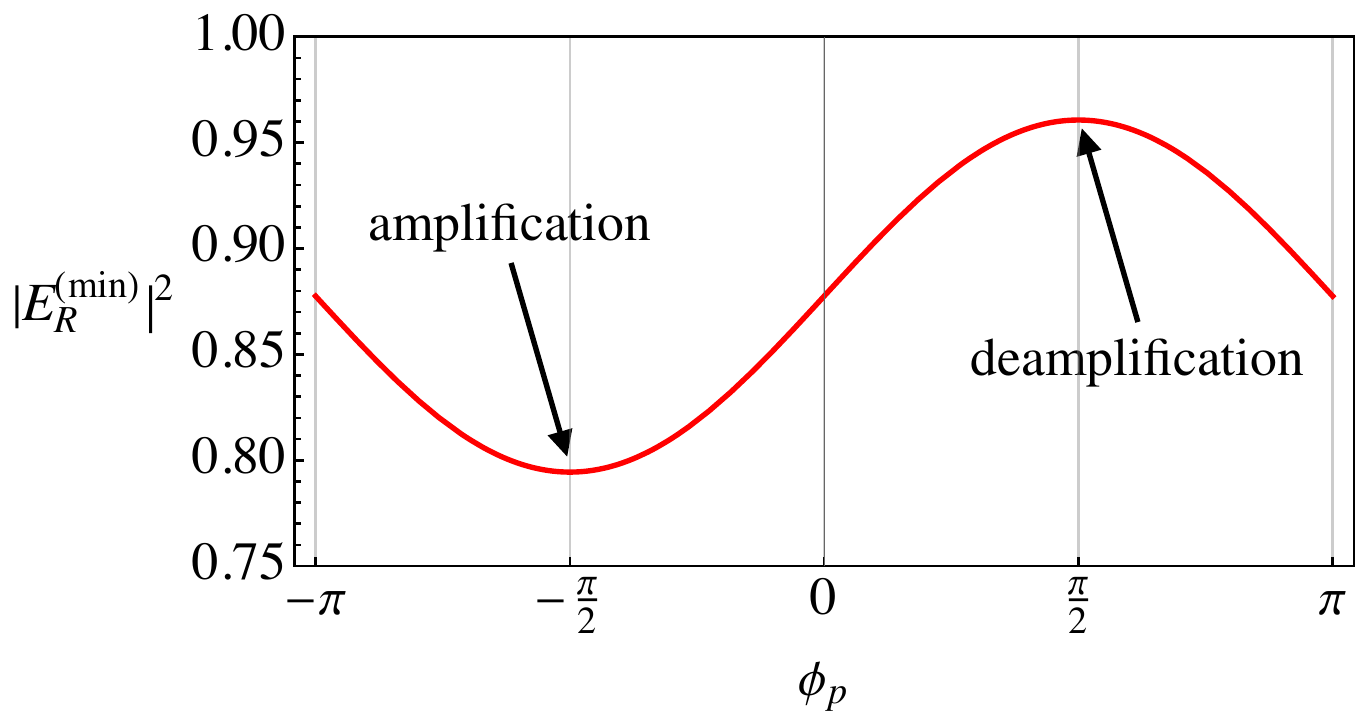}
\vspace{-0.2cm}
\caption{(Top and center) Plots of $|E_{\rm c}|^2$ and $|E_{R}|^2$ (dimensionless units) with $E_{\rm in}=1$ as functions of the phase $\phi$, proportional to the detuning $\nu$, and the pump phase $\phi_{\rm p}$. The blue lines refer to the behaviour of $|E_{\rm c}|^2$ and $|E_{R}|^2$ for a particular choice of the pump phase, form left to right $\phi_{p} = -\pi, -\pi/2, 0 , \pi/2$ and $\pi$.
(Bottom) Plot of the minimum value $|E_R^{\rm (min)}|^2$ of the reflected beam amplitude as a function of the pump phase $\phi_{\rm p}$. The other involved parameters have been set to the realistic values $R_1 = 0.999$, $R_2 = 0.9$, $\Delta = \SI{3.0e-3}{}$ and $|\gamma| = \SI{2.0e-2}{}$. See the text for details.
}
\label{fig:fields}
\label{fig:ERmin}
\end{figure}

In Fig.~\ref{fig:fields} (top and center panels) we plot $|E_{\rm c}|^2$ and $|E_{R}|^2$ as functions of the phase $\phi$, proportional to the detuning $\nu$, and of the pump phase $\phi_{\rm p}$ for realistic values of the other involved parameters (similar to the ones we shall use in our experiment described in Sect.~\ref{sec:results}). Remarkably, the minimum value of the intensity of the reflected beam, $|E_{R}^{\rm(min)}|^2$, depends on the pump phase $\phi_{\rm p}$ (see Fig.~\ref{fig:ERmin}, bottom panel). In particular, the minimum and the maximum of $|E_{R}^{\rm(min)}|^2$ occur in the amplification ($\phi_{\rm p}=-\pi/2$) and deamplification ($\phi_{\rm p}=\pi/2$) regimes, respectively, as shown in the bottom panel of Fig.~\ref{fig:ERmin}: between this two extremes, $|E_{R}^{\rm(min)}|^2$ is monotone with respect to $\phi_{\rm p}$ and this allows us to retrieve an \textit{error signal} to stabilize the pump phase. Note that, in other cavity configurations, namely, for a different choice of $R_1$ and $R_2$, one can find a \textit{minimum} of the reflected field and a \textit{maximum} of the transmitted one, but the effect of the $\phi_{\rm p}$ on their actual values is still the same. Therefore, also in this case we can retrieve the error signal by a suitable electronic inversion.

The stabilization of the cavity frequency is performed by the standard PDH technique, where a phase modulation $\phi_{\rm m}$ is added to the seed $E_{\rm in}$ at a frequency $\nu_m$ \cite{pdh}. 
The equation for the normalized error signal is (we recall that $\phi \propto \nu$) \cite{pdh,black}
\begin{equation}
\epsilon_{\rm PDH} = \Im \left[ E_R\left(\nu\right) E_R^* \left(\nu-\nu_m\right) -  E_R^*\left(\nu\right) E_R \left(\nu+\nu_m\right)  \right]\,.
\label{eq:err_pdh}
\end{equation}
\begin{figure}[tb]
\centering
\includegraphics[width=0.4\textwidth]{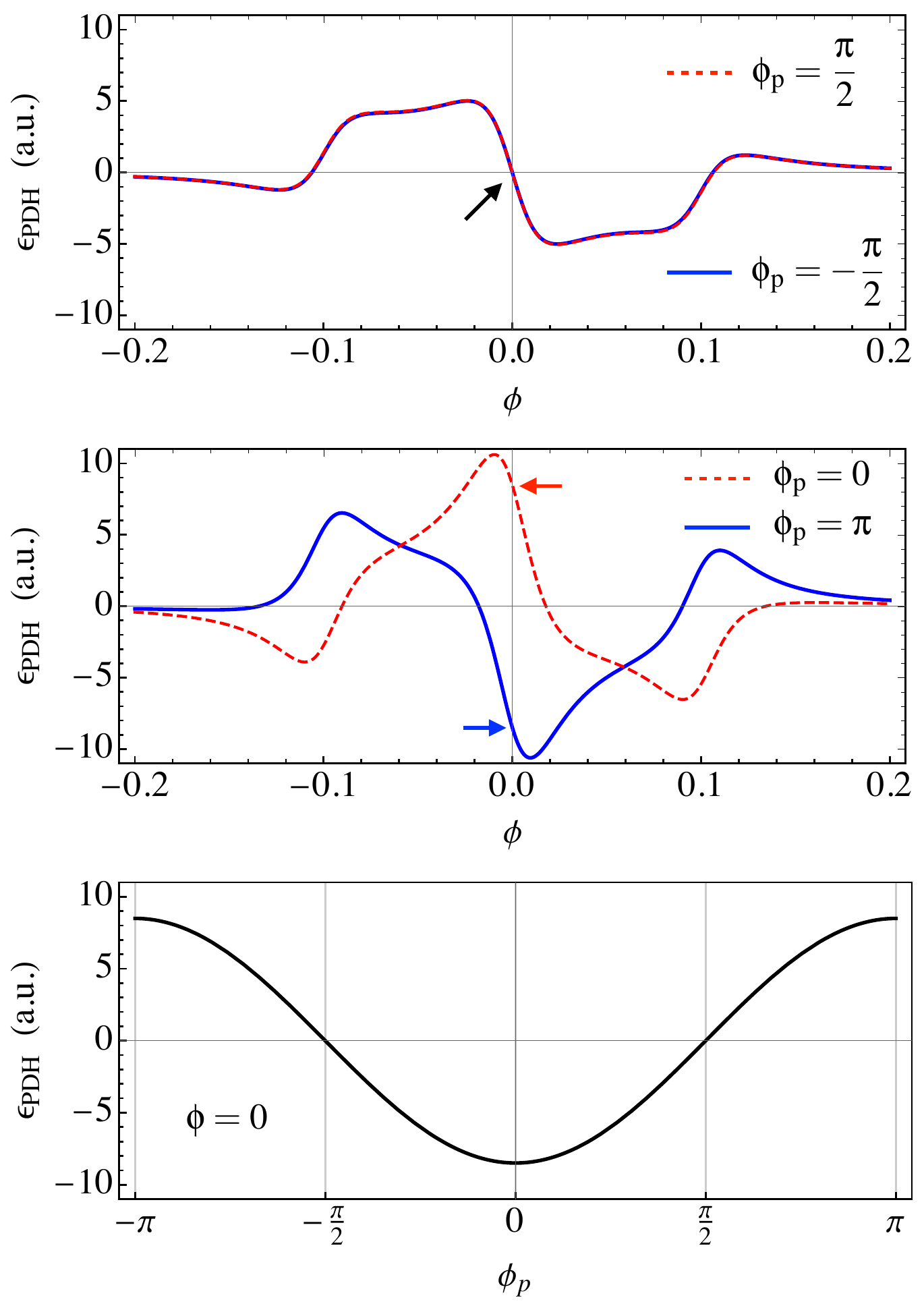}
\vspace{-0.4cm}
\caption{Plots of the PDH error signal $\epsilon_{\rm PDH}$ as a function of $\phi\propto \nu$ for fixed phase modulation $\phi_{\rm m} = 0.1$ and for different values of the pump phase $\phi_{\rm p}$. (Top) In the amplification ($\phi_{\rm p}=-\pi/2$) or in the deamplification ($\phi_{\rm p}=\pi/2$) regime the PDH error signal vanishes at resonance ($\phi = 0$). (Center) If $\phi_{\rm p}\ne \pm\pi/2$ the PDH error signal becomes zero at $\phi \ne 0$: there is an offset (arrows) due to the effect of the pump phase. (Bottom) Plot of $\epsilon_{\rm PDH}$ as a function of the pump phase $\phi_{\rm p}$ at resonance ($\phi=0$). The other involved parameters have been set to the realistic values $R_1 = 0.999$, $R_2 = 0.9$, $\Delta = \SI{3.0e-3}{}$ and $|\gamma| = \SI{2.0e-2}{}$.}
\label{fig:PDH:theory}
\end{figure}

\begin{figure}[tb]
\centering
\includegraphics[width=0.45\textwidth]{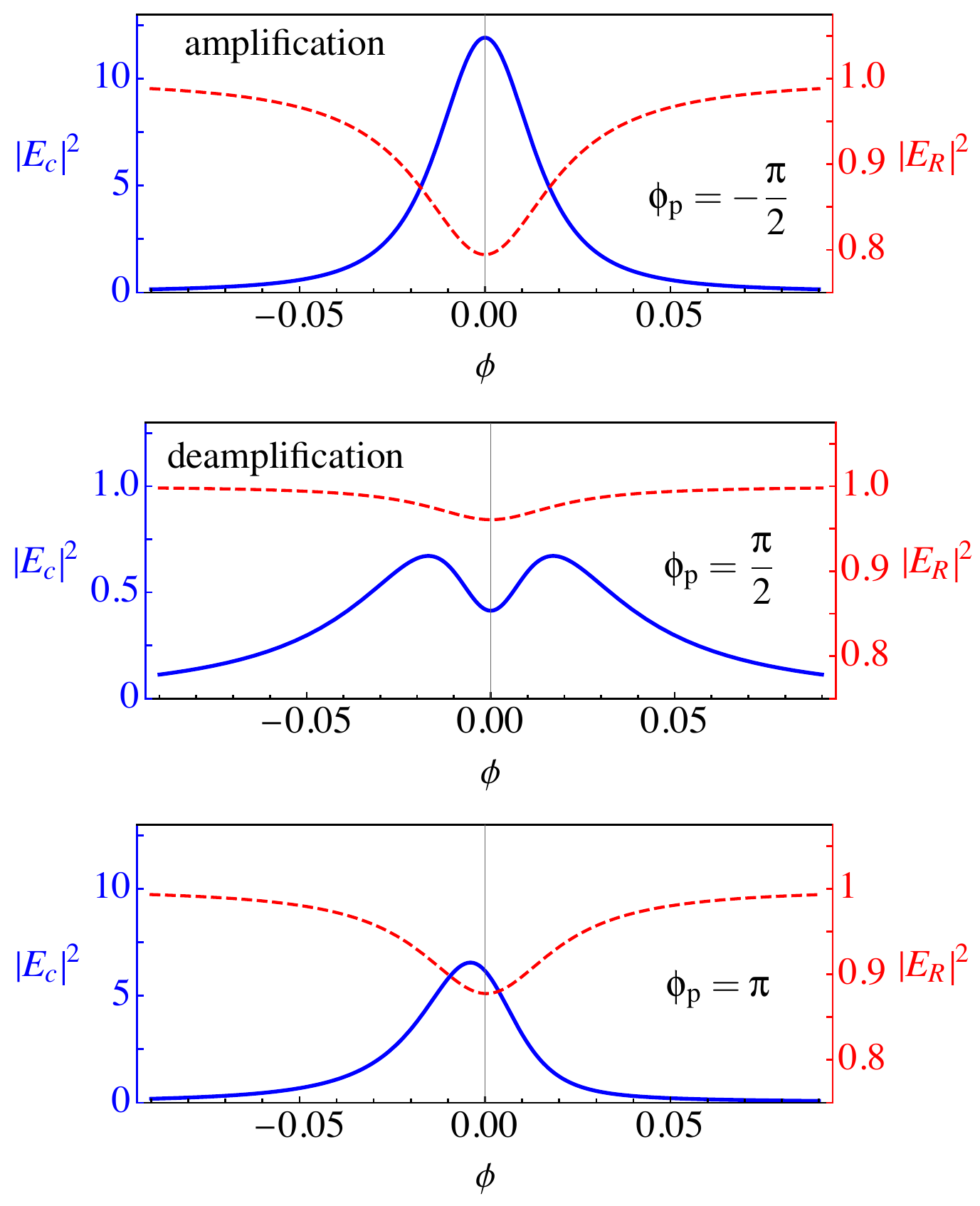}
\vspace{-0.4cm}
\caption{Plots of $|E_{\rm c}|^2$ (solid blue lines), proportional to the OPO output, and $|E_{R}|^2$ (dashed red lines) with $E_{\rm in}=1$ (dimensionless units) as functions of the phase $\phi \propto \nu$ for different values of the pump phase: (top) $\phi_{\rm p} = -\pi/2$, amplification regime; (center) $\phi_{\rm p} = \pi/2$, deamplification regime; (bottom) $\phi_{\rm p} = \pi$. The other involved parameters have been set to the realistic values $R_1 = 0.999$, $R_2 = 0.9$, $\Delta = \SI{3.0e-3}{}$ and $|\gamma| = \SI{2.0e-2}{}$.}
\label{fig:fields:res}
\end{figure}

Since $E_R$ depends on $\phi_{\rm p}$, the pump phase influences also the PDH error signal (\ref{eq:err_pdh}). This is a very important point and it will be better discussed in Sect.~\ref{sec:results}. Here we just observe that in the amplification or in the deamplification regime the PDH error signal vanishes at resonance, i.e. $\phi = 0$ (top panel of Fig.~\ref{fig:PDH:theory}): the field inside the cavity and, in turn, the transmitted beam, has a maximum or a minimum when the OPO is resonant with the input field, i.e. $\phi=0$, only in these two regimes (Fig.~\ref{fig:fields:res}, top and center panels). More in general, the pump phase $\phi_{\rm p}$ may lead to a PDH error signal which is no longer equal to zero at resonance (bottom panel of Fig.~\ref{fig:PDH:theory}). In order to have $\epsilon_{\rm PDH}=0$ also in this case,  the vertical offset has to be compensated. As mentioned, this offset vanishes in the amplification ($\phi_{\rm p}=-\pi/2$) or in the deamplification ($\phi_{\rm p}=\pi/2$) regime: the field inside the cavity and, in turn, the transmitted beam, has a maximum or a minimum when the OPO is resonant with the input field, i.e. $\phi=0$, only in these two regimes (Fig.~\ref{fig:fields:res}, top and center panels), otherwise the presence of the pump introduces a detuning of the resonance (Fig.~\ref{fig:fields:res}, bottom panel).
\begin{figure}[tb]
\centering
\includegraphics[width=0.4\textwidth]{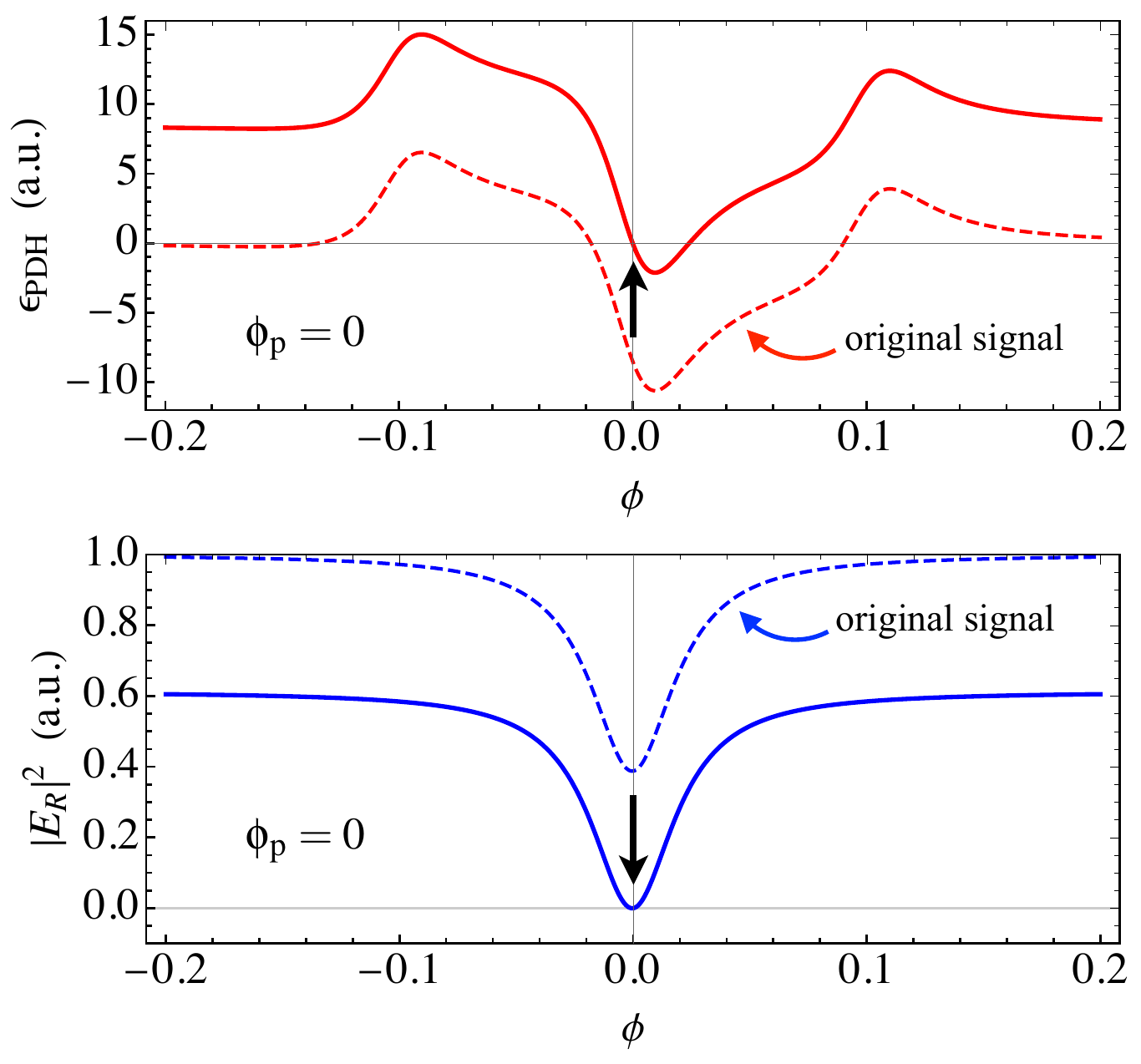}
\vspace{-0.4cm}
\caption{Plots of the PDH error signal $\epsilon_{\rm PDH}$  (top panel) and the amplitude of the reflected beam $|E_{R}|^2$ (bottom panel) as functions of $\phi \propto \nu$ for the pump phase $\phi_{\rm p} = \pi$. The dashed lines refer to the original signals (see the center panel of Fig.~\ref{fig:PDH:theory} and the lower panel of Fig.~\ref{fig:PDH:theory}), while the solid lines are obtained by setting to zero the $\epsilon_{\rm PDH}$ offset and the minimum of $|E_{R}|^2$, respectively. The other involved parameters have been set to the realistic values $R_1 = 0.999$, $R_2 = 0.9$, $\Delta = \SI{3.0e-3}{}$ and $|\gamma| = \SI{2.0e-2}{}$.}
\label{fig:fields:offsets}
\end{figure}
Overall, two different offsets have to be kept under control in order to generate two error signals: one for the OPO stabilization, obtained by centering the PDH error signal in zero (top panel of Fig.~\ref{fig:fields:offsets}), and one for seed-pump stabilizer (SPS), retrieved by fixing the minimum of the reflected beam in zero (bottom panel of Fig.~\ref{fig:fields:offsets}). In the next Sect.~\ref{sec:results} we shall show the experimental technique allowing to achieve both the goals.

\section{Experimental setup and results}\label{sec:results}
\begin{figure}[tb]
\centering
\includegraphics[width=0.45\textwidth]{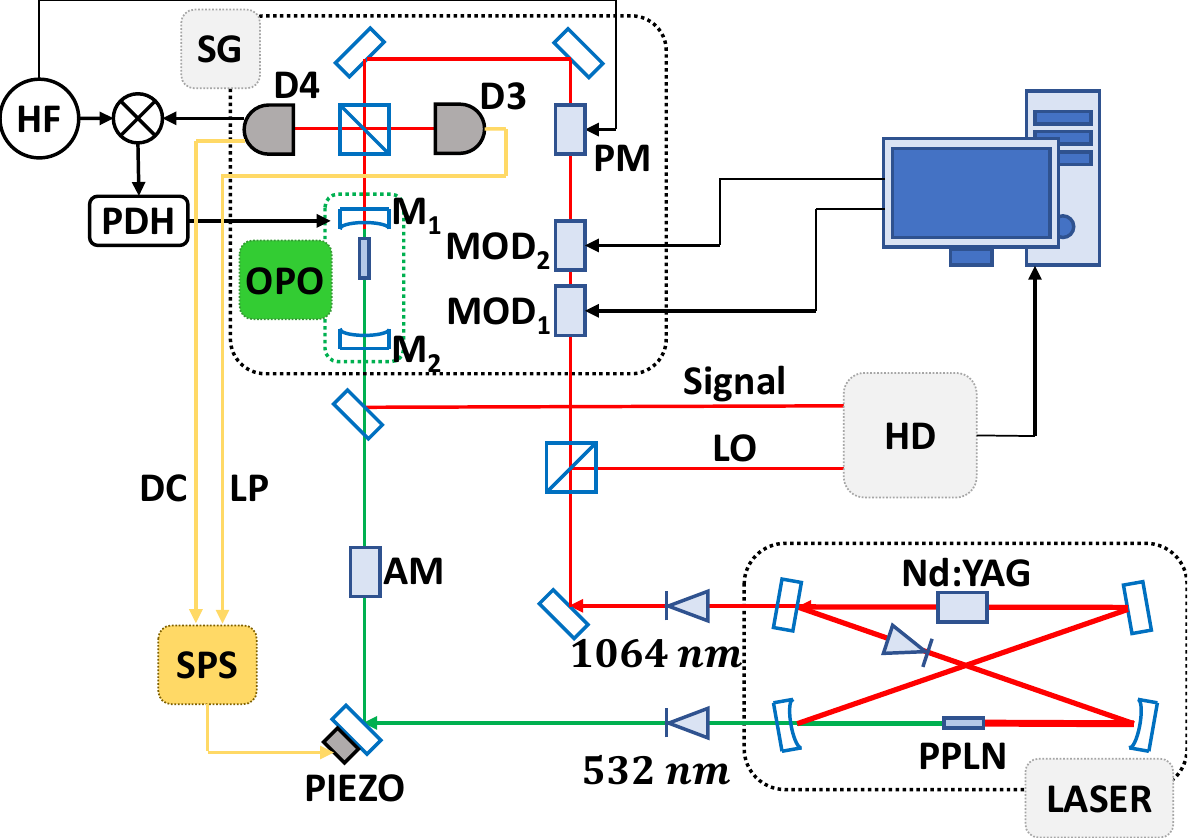}
\caption{Scheme of our setup.
The main parts are three: a LASER which produces both the pump @$\SI{532}{nm}$ and the seed and the local oscillator (LO) @$\SI{1064}{nm}$ for the state detection; a state generation part (SG) used for quantum states generation and squeezing; a detection part based on homodyne detection (HD). The OPO frequency is stabilized implementing the canonical PDH technique, while the seed-pump phase stabilization is performed by the seed-pump stabilizer (SPS) and is based on our technique. See the text for details.
}
\label{fig:setup}
\end{figure}

The theory introduced in the previous section has been tested using the experimental setup depicted in Fig.~\ref{fig:setup}.
Our setup allows us to generate, manipulate and detect displaced-squeezed states.
We have control on both the amplitude and the phase of the state and also on the amplitude and phase of the pump.
Following the scheme in Fig.~\ref{fig:setup}, the setup consists of three main blocks.
A homemade laser source (LASER) is a Nd:YAG and it generates both the seed (at \SI{1064}{\nano\meter}) and the pump (at \SI{532}{\nano\meter}) for the OPO, since it is internally frequency doubled with a PPNL crystal.
The internal second harmonic generation has a major advantage: it behaves as a damping force which suppresses the laser relaxing oscillations, dramatically decreasing its noise.
Therefore, this configuration avoids the necessity of building an external cavity for the second harmonic generation and of introducing other elements for the noise suppression \cite{WinP}.
The second block, the state generation (SG), is composed of two modulators ($\rm MOD_1$ and $\rm MOD_2$), which create quantum states on \SI{3}{\mega\hertz} sidebands, and an OPO used for squeezing.
States are then detected with a conventional homodyne detection scheme (HD), the third block of the setup, where a fraction of the source laser serves as the local oscillator (LO).

Both generation and detection are controlled by a computer.
The OPO is stabilized using a standard PDH technique \cite{pdh,black}, while the pump phase stabilization is performed with our method (SPS) based on the theoretical considerations given in Sect.~\ref{sec:theory}. As shown in Fig.~\ref{fig:setup}, the detector $\rm D3$ monitors the laser amplitude fluctuations (LP), so that the SPS error signal (DC) can be compensated for them. This error signal is manipulated with a PID and applied to a piezoelectric which variates the pump path lenght, resulting in a change of its phase in the OPO.
The laser has a free spectral range of \SI{200}{\mega\hertz}, while OPO of \SI{3025}{\mega\hertz}.

\begin{figure}[tb]
\centering
\includegraphics[width=0.235\textwidth]{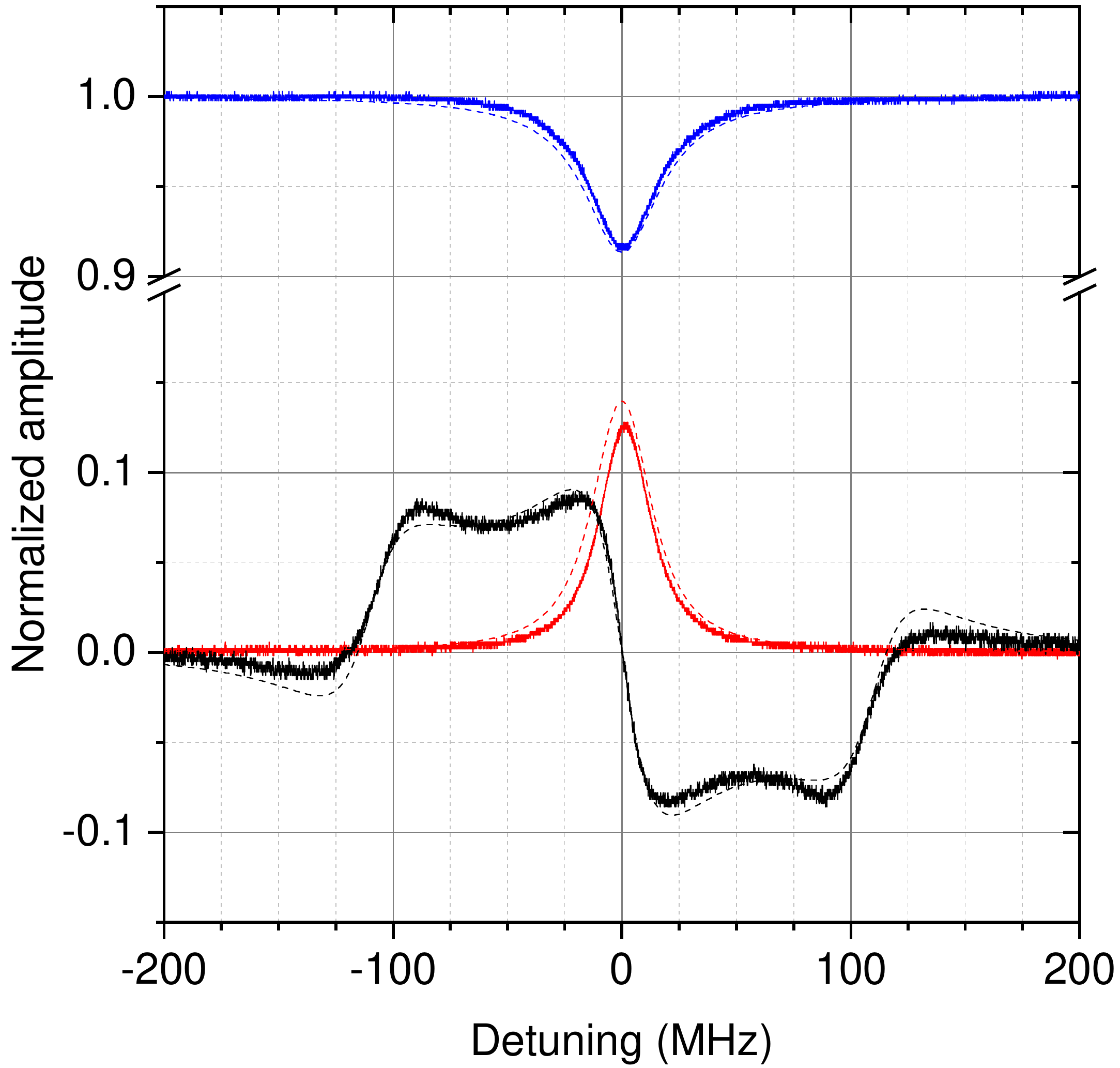}
\includegraphics[width=0.235\textwidth]{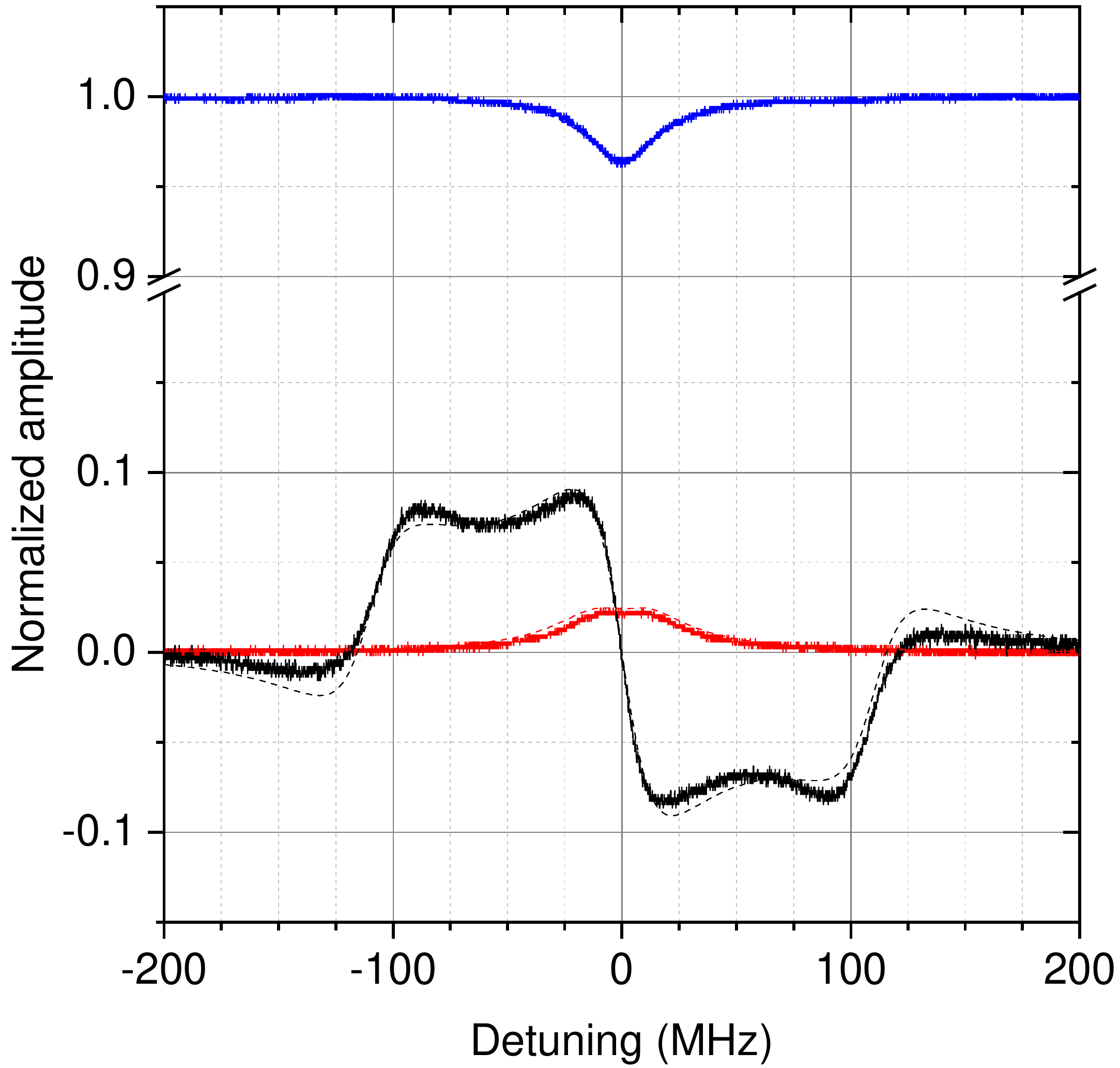}
\includegraphics[width=0.235\textwidth]{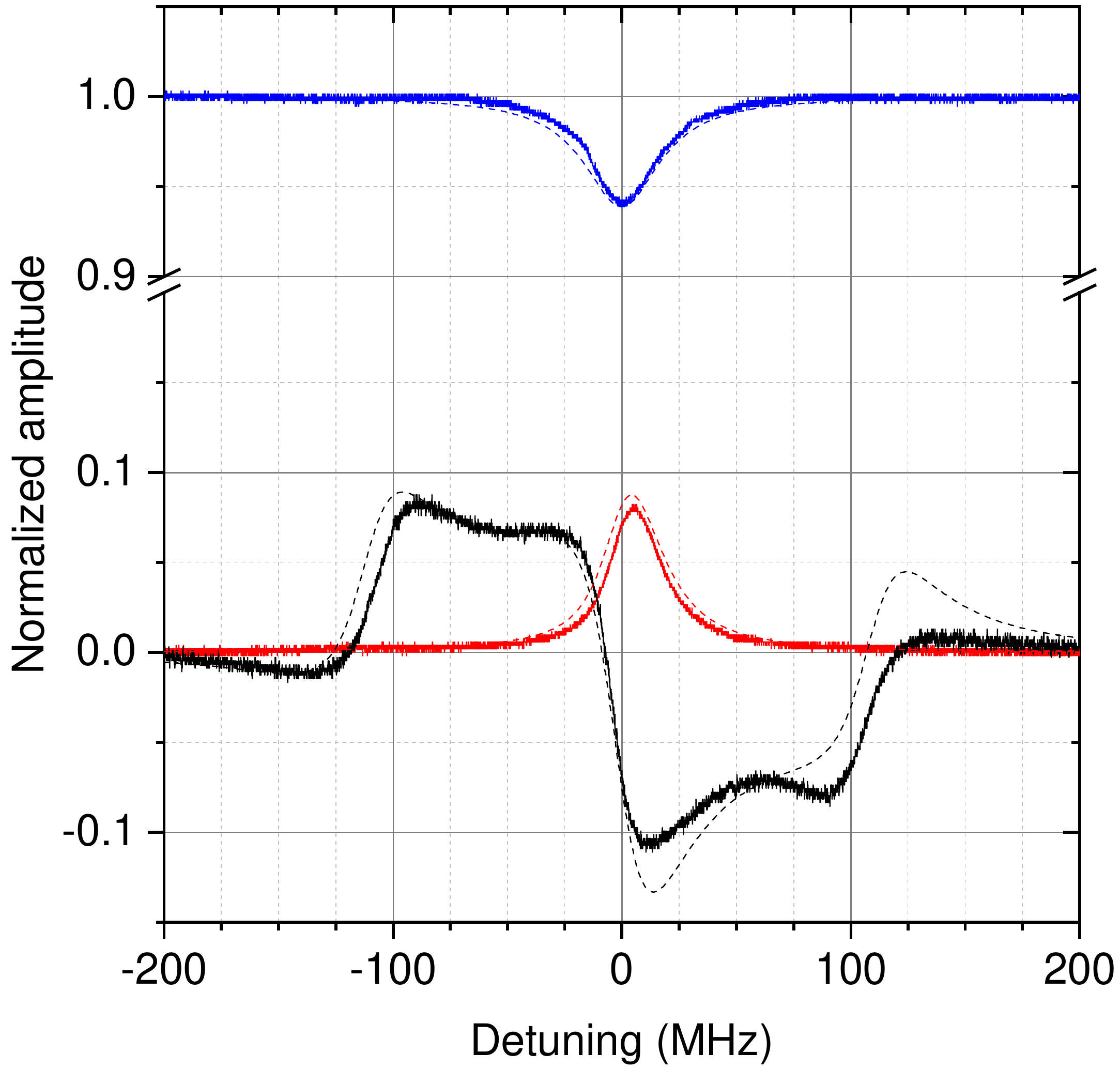}
\includegraphics[width=0.235\textwidth]{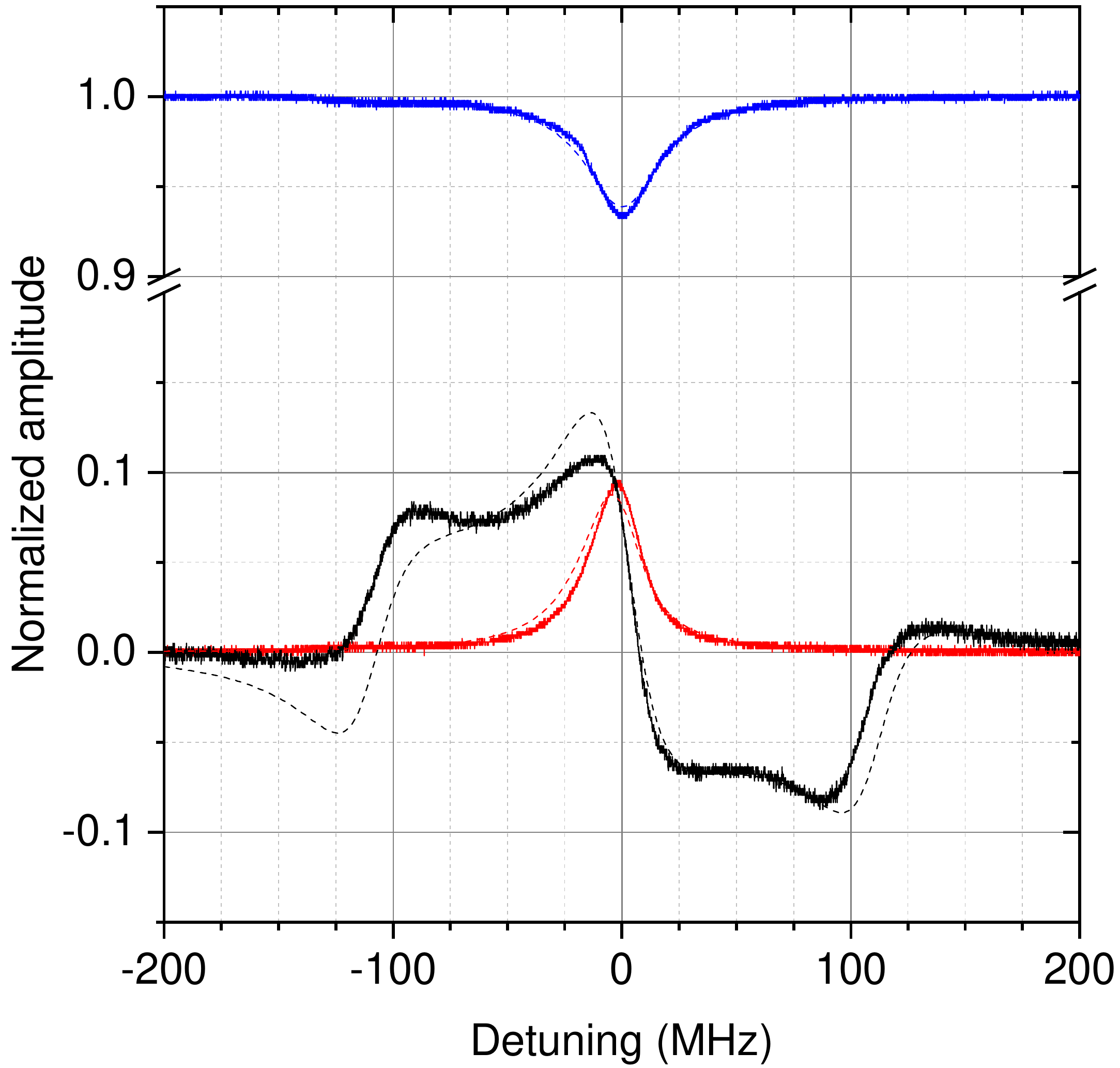}
\caption{Transmitted beam (red), reflected beam (blue) and PDH error signal (black) as functions of the detuning $\nu$, obtained with a frequency scan of the OPO cavity.
Solid lines are experimental data, while dashed lines are theoretical trends obtained from $R_1$, $R_2$, $\Delta$ and $\left|\gamma\right|$.
We plot curves in the amplification regime ($\phi_{\rm p}=-\pi/2$, upper left), deamplification regime ($\phi_{\rm p}=\pi/2$, upper right), \textit{minus} regime ($\phi_{\rm p}=0$, lower left) and \textit{plus} regime ($\phi_{\rm p}=\pi$, lower right). For the theoretical previsions we used the following measured values of the experimental parameters $R_1=0.9988$, $R_2=0.917$, $\Delta=\SI{2.4e-3}{}$ and $\left|\gamma\right|=\SI{1.85e-2}{}$.}
\label{fig:scan_opo}
\end{figure}
Figure~\ref{fig:scan_opo} shows the experimental intensity of the transmitted beam (red), of the reflected beam (blue) and of the PDH error signal (black) as functions of the detuning $\nu$, obtained with a scan of the OPO. The scan is performed by applying a linear voltage to a piezoelectric actuator attached to the output coupler of the OPO. The solid lines represents experimental data, while the dashed ones are the theoretical previsions.
In the figure we consider four configurations with different relative seed-pump phases.
The upper-left panel refers to the amplification regime, where $\phi_{\rm p}=-\pi/2$: in this configuration the transmitted beam is maximum, the reflected beam is minimum and the PDH error signal is zero when OPO is resonant with the laser.
The upper-right panel refers to the deamplification regime, where $\phi_{\rm p}=\pi/2$: as the previous case, both the transmitted and the reflected beams are centred around the OPO resonance, i.e. $\phi=\nu=0$, $\nu$ being the detuning. Notice the two small relative maxima in the transmitted beam around the resonance, both in theoretical and in experimental curves (see also the center panel of Fig.~\ref{fig:fields:res}).
The lower-left and lower-right panels of Fig.~\ref{fig:scan_opo} refer to two intermediate regimes that we call \textit{minus}, where $\phi_{\rm p}=0$, and \textit{plus}, respectively. We can clearly see that the PDH error signal has a negative (for $\phi_{\rm p}=0$) or positive ($\phi_{\rm p}=\pi$) offset for the OPO resonance (see also the center panel of Fig.~\ref{fig:PDH:theory}).

The experimental data are in very good agreement with the theoretical predictions obtained using the derived parameters of mirror reflectivities and non-linear crystal losses, namely, $R_1=0.9988$, $R_2=0.917$ and $\Delta=\SI{2.4e-3}{}$. Moreover, from the transmitted powers in amplification and deamplification regimes we calculated $G=5.68$ and finally derived $\left|\gamma\right|=\SI{1.85e-2}{}$ using Eq.~(\ref{eq:gain}).
We stress that in the two regimes \textit{minus} and \textit{plus} the PDH error signal presents an offset in the resonance condition, that is the error signal does not vanishes at resonance. If we shift the PHD signal offset to zero with a suitable compensation, the PDH technique will allow to stabilize the OPO cavity at resonance also with $\phi_{\rm p} \ne \pm \pi/2$. Moreover, since the actual value of the minimum of the reflected beams is a monotone function of the pump phase for
$(2k-1)\pi/2+k\pi < \phi_{\rm p} < (2k+1)\pi/2$, $k \in {\mathbbm Z}$, as one can see from Fig.~\ref{fig:ERmin}, we can always set this minimum to zero in order to retrieve an error signal to stabilize the pump phase. This point will be better clarified in the following.

\begin{figure}[tb]
\centering
\includegraphics[width=0.3\textwidth]{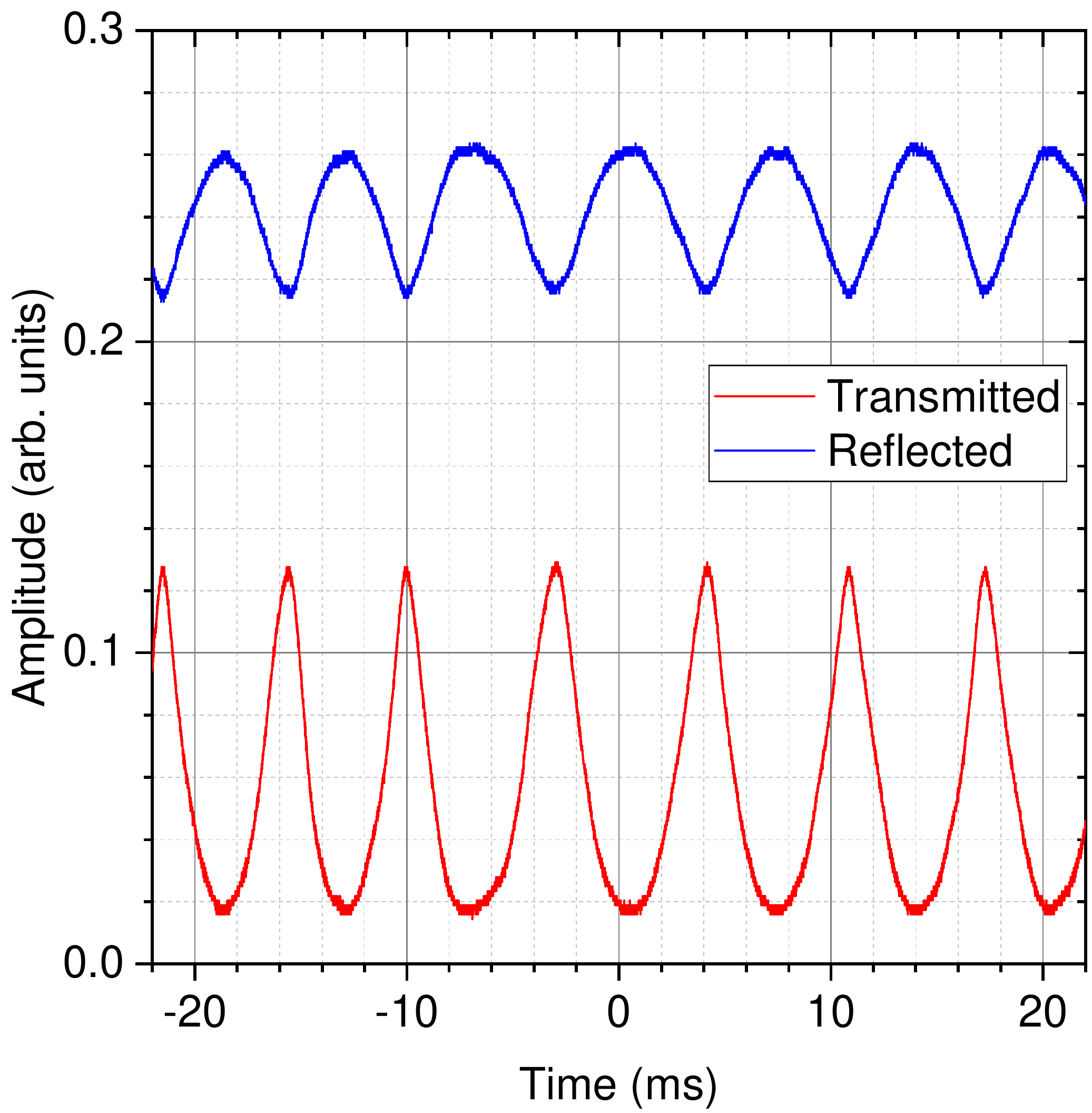}
\vspace{-0.2cm}
\caption{Transmitted beam (red) and reflected beam (blue) for a scan of the cavity (here the time is related to the pump phase $\phi_{\rm p}$). A triangular signal has been applied to the PIEZO in Fig.~\ref{fig:setup} acting on the pump phase.}
\label{fig:scan_pump}
\end{figure}

Figure~\ref{fig:scan_pump} shows the amplitudes of the reflected and transmitted beams for a pump phase scan.
A piezoelectric actuator is attached to a mirror used to change the pump phase $\phi_{\rm p}$ and a linear voltage is applied.
The OPO is actively stabilized during the scan, where we put the PDH error signal offset to zero.
It is clear that the reflected beam has a maximum in the deamplification regime and a minimum in the amplification regime, as expected.
Thus, after the application of a proper offset, the reflected beam power is a good error signal and can be used to stabilize $\phi_{\rm p}$ at any value but $\pm\pi/2$, where the derivative of the error signal vanishes.

The advantage of our technique is that both PDH error signal for OPO stabilization and SPS error signal for the pump phase stabilization are obtained from the same reflected beam.
This allows us to not touch the transmitted beam, which cannot suffer power losses since it contains the squeezed state.

\begin{figure}[tb]
\centering
\includegraphics[width=0.312\textwidth]{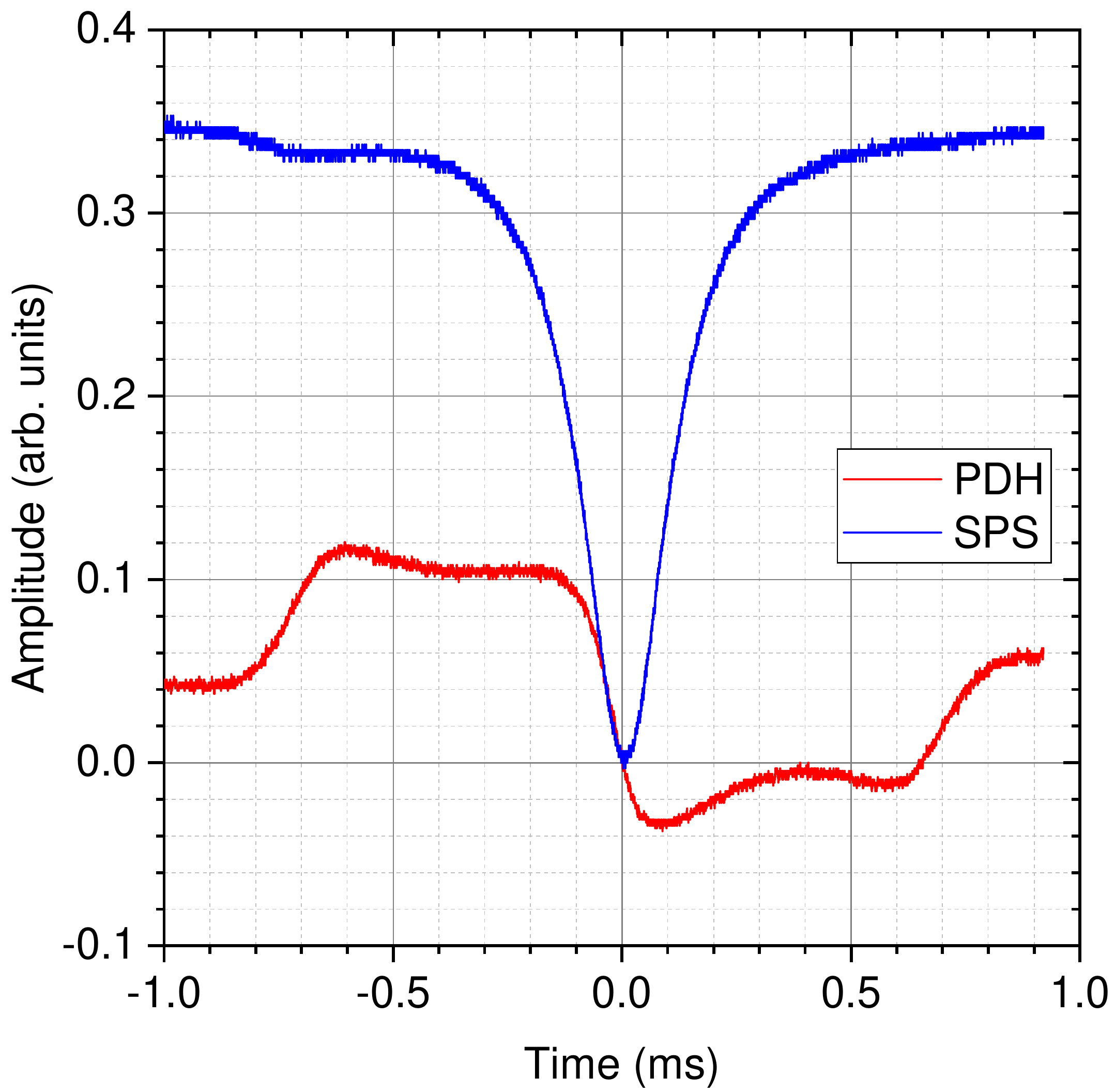}
\vspace{-0.2cm}
\caption{PDH error signal (red) and SPS error signal (blue) during a scan of the cavity frequency with all the offsets optimized (here the time is related to the phase $\phi \propto \nu$).}
\label{fig:opt_err}
\end{figure}

The procedure for the optimization of the two offsets (PDH and SPS) is rather simple.
Once the phase of the pump is chosen and, thus, the amplitude of the reflected beam at resonance, a scan of the OPO is performed and the two error signals mentioned above are monitored. An offset is added electronically to the PDH error signal in order to set it to zero where the reflected beam has its minimum, while another offset is added to the SPS error signal to shift to  zero the value of the maplitude of the reflected beam at the same point (see also Fig.~\ref{fig:fields:offsets}).
Figure~\ref{fig:opt_err} shows the two experimental signals after this optimization in a configuration near the \textit{minus} configuration.
\begin{figure}[tb]
\centering
\includegraphics[width=0.3\textwidth]{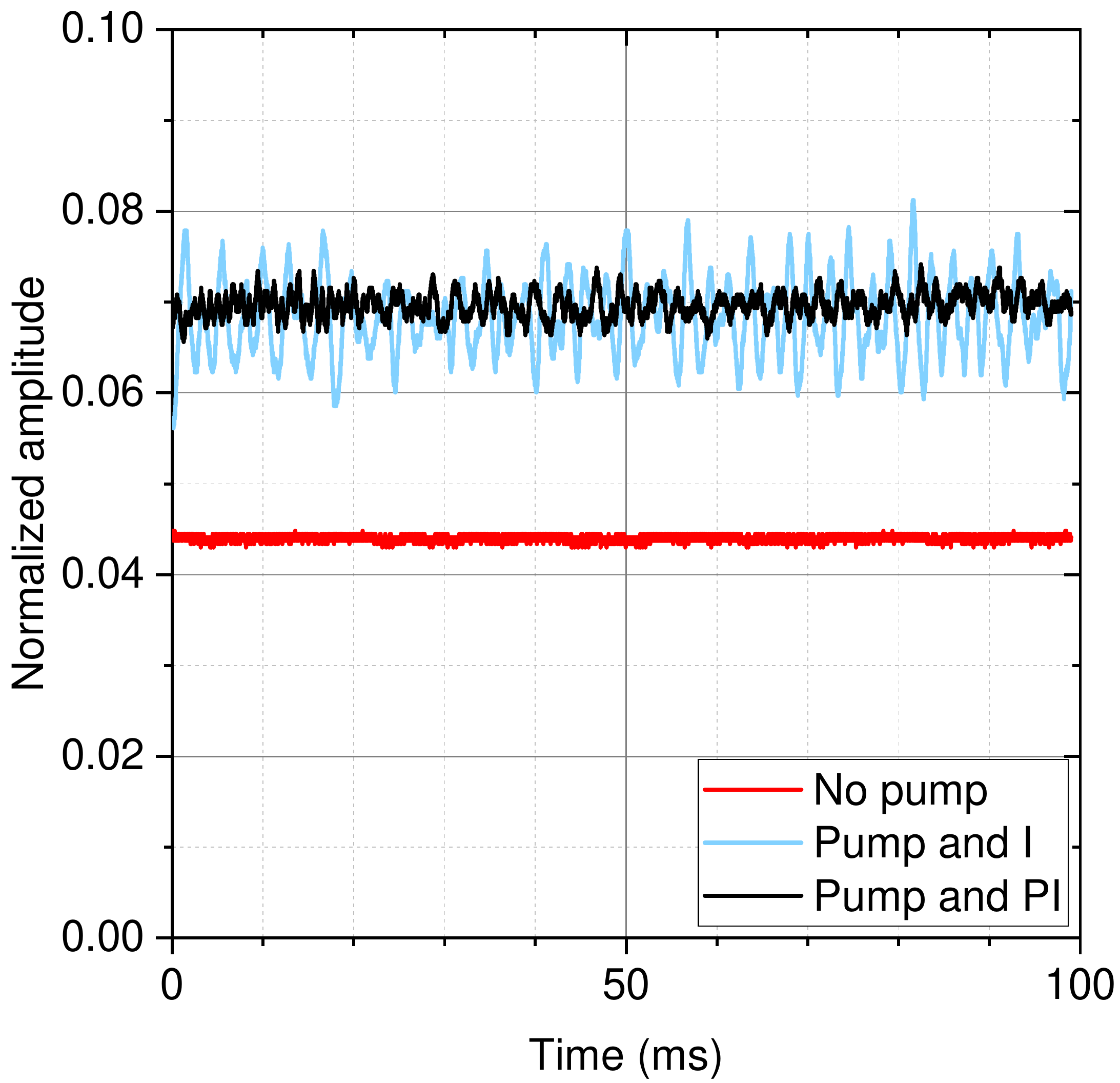}
\vspace{-0.2cm}
\caption{Example of OPO stabilization in different conditions as a function of time. Lines represent the transmitted beam in three cases: pump turned off (red), pump on and pump phase stabilized at $\phi_{\rm p}=0$ with both integrator and proportional (black) and pump on and pump phase stabilized at $\phi_{\rm p}=0$ with the only integrator (light blue). The relative intensity noise is \SI{0.6}{\percent}, \SI{1.9}{\percent} and \SI{6.2}{\percent}, respectively.
Notice that the black line has been rigidly shifted of -0.4 for a better visualization.}
\label{fig:stab}
\end{figure}
In order to compensate for the SPS error signal for the laser power fluctuations, we subtract the direct laser power from it.
Finally, both the resulting SPS error signal and the PDH error signal are independently processed with two homemade PIDs and applied to the piezoelectric actuator for changing $\phi_{\rm p}$ and the cavity length, respectively.
Precisely, our PID is actually a PI, with an integrative and a proportional part.
Figure~\ref{fig:stab} displays the amplitude of the transmitted beam of the OPO in three different cases.
The red lines refers to the condition without the pump and the relative power fluctuations are \SI{0.6}{\percent}.
The black lines refers to the \textit{minus} condition ($\phi_{\rm p}=0$). Both the integrative and the proportional of the SPS PID are activated and the relative power fluctuations are \SI{1.9}{\percent}.
Finally, the blue lines refers again to the \textit{minus} condition, but now the proportional is switched off and the relative power fluctuations are \SI{6.2}{\percent}. In this last case the fluctuations have the typical frequency of the mountings mechanical vibrations, which is of the order of \SI{}{\kilo\hertz}.
Black and blue curves have been rigidly shifted in the amplitude direction of $-0.2$ and $-0.4$, respectively, in order to better visualize them.
It is worth noting that, once the OPO and the pump phase are stabilized, the system remains stable for a long time, even hours. The only limiting factor is the dynamics of the two piezoelectric actuators, which can reach their maximum (or minimum) elongation if the thermal deformations are too large.

\section{Application: reliable generation of displaced squeezed states}\label{s:application}
\begin{figure}[tb]
\centering
\includegraphics[width=0.45\textwidth]{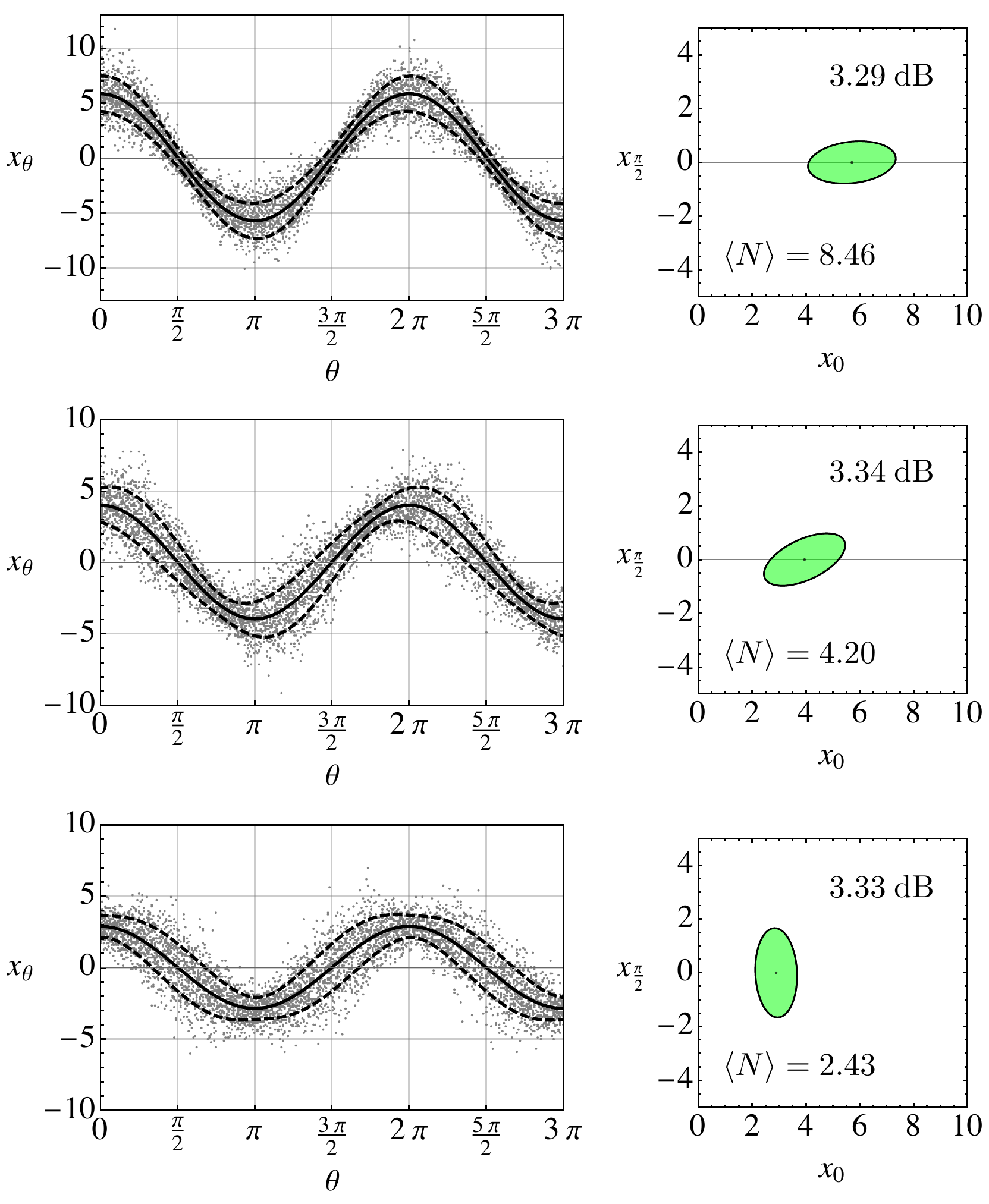}
\vspace{-0.2cm}
\caption{Examples of displaced squeezing states generation with different: (left plots) homodyne traces and the corresponding tomographic reconstruction in the phase space (right plots).
(Top) Amplification regime with $\phi_{\rm p} \approx -\pi/2 \Rightarrow \arg(\xi) \approx 0$, $\alpha = 2.93$ and $N_{\rm th}=0.13$;
(Center) $\phi_{\rm p} \approx 0 \Rightarrow \arg(\xi) \approx \pi/2 $, $\alpha = 1.97$ and $N_{\rm th}=0.12$;
(Bottom) deamplification regime $\phi_{\rm p} \approx \pi/2 \Rightarrow \arg(\xi)\approx \pi$, $\alpha = 1.44$ and $N_{\rm th}=0.14$.
In the left plots, the solid lines refer to the average value of the experimental trace, while the dashed lines represents the standard deviation, better highlighting the effect of squeezing.
Note that, in the phase space (right plots), the angle between the major axis of the ellipses and the horizontal one is given by $\arg(\xi)/2$. In the pictures we also report the corresponding squeezing level and total average number of photons. The vacuum variance is set to 1.
}
\label{fig:squeezed_states}
\end{figure}
In this Section we demonstrate the potentiality of our technique by applying the method to create displaced squeezed states. More precisely, taking into account the losses, the density operator describing the single-mode states we generate and anlayze can be written in the following compact form \cite{oli:rev}:
\begin{equation}\label{gen:state}
\rho = D(\alpha) S(\xi) \rho_{\rm th} S^\dag(\xi) D^\dag(\alpha)
\end{equation}
where, as usual,  $D(\alpha) = \exp(\alpha a^\dag - \alpha^* a)$ and
$S(\alpha) = \exp[\frac12\xi a^2 - \frac12 \xi^* (a^\dag)^2]$ are the displacement and squeezing operators, respectively, $a$ being the annihilation operator of the field, $[a,a^\dag] = {\mathbbm I}$. In the previous formula we also introduced the density operator of the {\it thermal state}
\begin{equation}
\rho_{\rm th} = \sum_{n=0}^{\infty} \frac{(N_{\rm th})^n}{(1+N_{\rm th})^{n+1}}|n\rangle \langle n|\,.
\end{equation}
We remark that the state (\ref{gen:state}) is the {\it overall} state generated by our setup \cite{oli:rev, notarnicola}. In this context, the average number of thermal photons $N_{\rm th}$ can be seen as an effective parameter summarizing the effect of losses (only if $N_{\rm th}=0$ we have no losses at all and the generated state is thus pure). For such a state the squeezing level in dB is given by $-10 \log_{10}\left[(1+2N_{\rm th} ) e^{-2|\xi|}\right]$.
For the sake of simplicity, but without loss of generality, we considered states with $\alpha \in {\mathbbm R}$ and $\xi \in {\mathbbm C}$ and we also obtain $\arg(\xi) = \phi_{\rm p} +\pi/2$.

The addressed states have been generated through our setup following the well-established procedure described in Refs.~\cite{cialdi,olivares,porto,mandarino}. Figure~\ref{fig:squeezed_states} shows the experimental homodyne traces of three examples of different states (left) and their tomographic reconstructions, represented in the phase space (right). In the plots it is clear the modulation of the quadrature standard deviation due to the squeezing (dashed lines in the left plots) as well as the relative phase between the coherent amplitude (set to zero) and the squeezing parameter $\xi$  (related to the angle between the $x_0$-axis and the major axis of the ellipses in the right plots). The latter is indeed a proof of the reliability of our pump-seed phase stabilization technique, being the pump and seed phases connected to the the squeezing parameter and coherent amplitude ones.

We remark that, since the aim of our study was to demonstrate the effectiveness of the technique we developed, in our experiments we used a pump laser with a not very high intensity, thus resulting in a squeezing level around \SI{3}{\decibel} as shown in Fig.~\ref{fig:squeezed_states}.

\section{Conclusions}\label{s:concl}
In this work we presented the theoretical model and the experimental verification of an innovative method for the stabilization of the relative phase between seed and pump of an OPO. To this aim, we have also developed a proper model of the OPO to obtain the analytical expression fo the amplitude of the beam reflected off the cavity as a function of the cavity frequency, of the crystal nonlinearity and of the pump phase. In particular, our technique allows to extract two error signals from for the reflected beam, one for the stabilization of the pump-seed phase and one the OPO frequency with PDH technique, without the necessity of two different modulation and demodulation systems required by other methods (see, for instance, Appendix~\ref{alternative}). Our analysis shows that the pump affects the PDH error signal adding an offset that has to be compensated and that depends on the pump phase. We have  also shown that our system allows the suppression of the noise caused by pump phase fluctuations to \SI{1.9}{\percent} on the transmitted beam, leading to an improvement of the squeezing level. Eventually, the reliability of our technique can pave the way not only for practical applications exploitingn squeezed coherent states, but also to study more fundamental aspects of quantum information science based on continuous-variable systems.

\section*{Acknowledgments}
This work has been supported by MAECI, Project No.~PGR06314 ``ENYGMA''.

\appendix
\section{Alternative method for pump stabilization}\label{alternative}
In this Appendix we present another method for the pump stabilization that is suggested in \cite{bowen}: it is based on modulating the pump entering the OPO with sidebands. In this case the $\gamma$ parameter can be written as
\begin{equation}
\gamma_m =  \gamma - i \beta \sin(2\pi \nu_m t).
\end{equation}
The two sidebands of the pump stimulates the generation of two sideband at frequency $\nu_m$, which are not resonant with the OPO cavity.
This leads to an error signal with three terms: the central term $E_R$ and the two sidebands at frequency $\nu_m$ given by the transmitted part of $E_R$ on the sidebands.
Thus the error signal for this technique can be written as
\begin{equation*}
\epsilon_{\rm B} =  \Im \left[ E_R(\nu) E_m^* (\nu) -  E_R^*(\nu)  E_m (\nu)  \right]
\label{eq:err_B}
\end{equation*}
where the field $E_m$ can be obtained from $E_R$ in Eq.~(\ref{eq:Er_final}) by taking into account the contribution from $i \gamma (1-R_1)$. Therefore, one has
\begin{equation}
E_m = i \gamma \frac{\sqrt{1-R_1}}{\sqrt{R_1} } E_c^*.
\label{eq:em}
\end{equation}
It is clear that, in order to work, the present technique requires a modulation on the pump as well as a modulation on the seed for the PDH OPO stabilization. This last modulation has to be performed at a different frequency with respect to the pump modulation. For this reason, one needs two different modulation, demodulation and detection stages, which is a possibile disadvantage.


\begin{thebibliography}{50}

\bibitem{GQI} C. Weedbrook, S. Pirandola, R. Garc\'ia-Patr\'on, N. J. Cerf, T. C. Ralph, J. H. Shapiro, and S. Lloyd, {\it Gaussian quantum information}, Rev. Mod. Phys. {\bf 84}, 621--669 (2012).

\bibitem{lough} J. Lough, E. Schreiber, F. Bergamin, H. Grote, M. Mehmet, H. Vahlbruch, C. Affeldt, M. Brinkmann, A. Bisht, V. Kringel, H. L\"uck, N. M., S. Nadji, B. Sorazu, K. Strain, M. Weinert, and K. Danzmann, {\it First demonstration of 6 dB quantum noise reduction in a kilometer scale gravitational wave observatory}, Phys. Rev. Lett. {\bf 126}, 041102 (2021).

\bibitem{breite} G. Breitenbach, S. Schiller, and J. Mlynek, {\it Measurement of the quantum states of squeezed light}, Nature {\bf 387}, 471--475 (1997).

\bibitem{dauria09} V. D'Auria, S. Fornaro, A. Porzio, S. Solimeno, S. Olivares, and M. G. A. Paris, {\it  Full characterization of Gaussian bipartite entangled states by a single homodyne detector}, Phys. Rev. Lett. {\bf 102}, 020502 (2009).

\bibitem{buono12} D. Buono, G. Nocerino, V. D'Auria, A. Porzio, S. Olivares, and M. G. A. Paris, {\it Quantum characterization of bipartite Gaussian states}, J. Opt. Soc. Am. B {\bf 27}, A110--A118 (2010).

\bibitem{porto} C. Porto, D. Rusca, S. Cialdi, A. Crespi, R. Osellame, D. Tamascelli, S. Olivares, and Matteo G. A. Paris, {\it Detection of squeezed light with glass-integrated technology embedded into a homodyne detector setup},
J. Opt. Soc. of Am. B  {\bf 35}, 1596--1602 (2018).

\bibitem{mandarino} A. Mandarino, M. Bina, C. Porto, S. Cialdi, S. Olivares, and M. G. A. Paris,  {\it Assessing the significance of fidelity as a figure of merit in quantum state reconstruction of discrete and continuous-variable systems}, Phys. Rev. A  {\bf 93}, 062118 (2016).

\bibitem{pdh} R. Drever, J. Hall, F. Kowalski, J. Hough, G. Ford, A. Munley, and H. Ward,
{\it Laser phase and frequency stabilization using an optical resonator},
Appl. Phys. B {\bf 31}, 97--105 (1983).

\bibitem{heurs} M. Heurs, I. R. Petersen, M. R. James, and E. H. Huntington, 
{\it Homodyne locking of a squeezer}, Opt. Lett. {\bf 34}, 2465--2467 (2009).

\bibitem{hansch} T. Hansch, and B. Couillaud, {\it Laser frequency stabilization by polarization spectroscopy of a reflecting reference cavity}, Opt. Comm. {\bf 35}, 441--444 (1980).

\bibitem{robins} N. P. Robins, B. J. J. Slagmolen, D. A. Shaddock, J. D. Close, and M. B. Gray, {\it Interferometric, modulation-free laser stabilization}, Opt. Lett. {\bf 27}, 1905--1907 (2002).

\bibitem{bowen} W. P. Bowen,  {\it Experiments towards a quantum information network with squeezed light and entanglement}, PhD Thesis,  School of Physical Sciences, Australian National University (2003).

\bibitem{sun} X. Sun, Y. Wang, L. Tian, S. Shi, Y. Zheng, and K. Peng,  {\it Dependence of the squeezing and anti-squeezing factors of bright squeezed light on the seed beam power and pump beam noise}, Opt. Lett. {\bf 44}, 1789--1792 (2019).

\bibitem{schnabel} H. Vahlbruch, S. Chelkowski, B. Hage, A. Franzen, K. Danzmann, and R. Schnabel,
{\it Coherent control of vacuum squeezing in the gravitational-wave detection band}, Phys. Rev. Lett. {\bf 97}, 011101 (2006).

\bibitem{denker} T. Denker, D. Schutte,  M.  H. Wimmer, T. A. Wheatley, E. H. Huntington, and M. Heurs
{\it Utilizing weak pump depletion to stabilize squeezed vacuum states}, Opt. Exp. {\bf 23}, 16517--16528 (2015).

\bibitem{yariv} A. Yariv, and P. Yeh, {\it Optical Wave in Crystals: Propagation and Control of Laser Radiation} (Wiley-Interscience, 2002).

\bibitem{black} E. D. Black, {\it An introduction to Pound--Drever--Hall laser frequency stabilization}, Am. J. Phys. {\bf 69}, 79--87 (2001)

\bibitem{WinP} S. Cialdi, and E. Suerra, in preparation.

\bibitem{oli:rev}S. Olivares, {\it Quantum optics in the phase space}, Eur. Phys. J. Special Topics {\bf 203}, 3--24 (2012).

\bibitem{cialdi} S.Cialdi, E. Suerra, S. Olivares, S. Capra, and M. G. A. Paris, {\it Squeezing phase diffusion},
Phys. Rev. Lett. {\bf 124}, 163601 (2020).

\bibitem{olivares} S. Olivares, S. Cialdi, and M. G. A. Paris, {\it Homodyning the $g^{2}(0)$ of Gaussian states},
Opt. Comm. {\bf 426}, 547--552 (2018).

\bibitem{notarnicola} M. N. Notarnicola, M. G. Genoni, S. Cialdi, M. G. A. Paris, and S. Olivares, {\it Phase noise mitigation by realistic optical parametric oscillator}, arXiv:2106:11631 [quant-ph].

\end{thebibliography}
\end{document}